\documentclass{elsart}
\usepackage[dvips,dvipdfm]{graphicx, color}
\usepackage{amsmath}
\usepackage{psfrag}
\usepackage{amsfonts}
\usepackage{amssymb}

\begin{document}
\begin{frontmatter}
\title{Pulse Dynamics in Coupled Excitable Fibers: Soliton-like Collision, Recombination, and Overtaking}

\author{Hiromichi Suetani$^{\rm a, c}$, Tatsuo Yanagita$^{\rm b}$, and Kazuyuki Aihara$^{\rm c, a}$}
\address[Sue]{Aihara Complexity Modelling Project, ERATO, JST, Tokyo 151-0064, Japan}
\address[Yan]{Research Institute for Electronic Science, Hokkaido University, Sapporo 060-0812, Japan}
\address[Aih]{Institute of Industrial Science, The University of Tokyo, Tokyo 153-8505, Japan}

\date{\today}

\begin{abstract}
We study the dynamics of a reaction-diffusion system composed of two mutually coupled excitable fibers. 
We focus on the situation in which dynamical properties of the two fibers are not identical  because of  the parameter difference between the fibers. 
Using the spatially one-dimensional FitzHugh-Nagumo equations as a model of a single excitable fiber, we show that the system exhibits a rich variety of dynamical behavior, including soliton-like collision between two pulses, recombination of a solitary pulse and  synchronized pulses, and overtaking of a slow-moving solitary pulse by fast-moving synchronized pulses.
\end{abstract}



\end{frontmatter}

\section{Introduction}
Excitability is a ubiquitous dynamical property encountered in many fields of science and plays  important roles in the functional aspects of many living systems such as transmission of electronic signals in neural and cardiac systems~\cite{Murr89,Keen98,Winf01}. 
It has been found that spatially extended excitable media, which are modeled  in the framework of reaction-diffusion systems,  
show a rich variety of dynamical behavior including propagating pulses and  target waves~\cite{Zhab70}, spiral waves~\cite{Winf72}, and spatio-temporal chaos~\cite{Bar93}. 

Problems resulting from {\it intra-medium} interactions of  spatially localized patterns in an excitable medium have attracted great interest. 
One of the distinguishing features of such an interaction is that two propagating pulses in an excitable medium annihilate each other upon head-on collision~\cite{Tasa49}. 
In general, dissipative systems feature not only annihilation of excited waves; they have a variety of interactions among spatially localized patterns that behave like an elastic object upon collision and scatter in various ways~\cite{Tuck78,Bar92,Petr94,Kris94,Kose95,Ohta97,Arge97,Nish03}.  
%

%
On the other hand, {\it inter-media} interactions, i.e., interactions among spatially localized patterns in multilayered  excitable media, should be also of great importance from the practical viewpoint.  
%
%
%
For example,  in several nerve systems such as the hippocampus, olfactory nerves, corpus callosum, spinal column, peripheral nerves, and cerebellum, it is observed that huge nerve axons are arranged in densely packed bundles so that neighboring neurons can electronically communicate with each other. 
Beginning  with  the pioneering works of Katz and Schmitt~\cite{Katz40} and Arvanitaki~\cite{Arva42}, studies on electrical axo-axonal interactions have had a long history~\cite{Ramo78,Selt79,Rasm80,Kocs82,Meye85,Boki01,Blin03}. 
Besides nerve systems, similar bundled structures are observed in cardiac systems such as the bundle of His and Purkinje fibers in the  myocardium. 
%
Mathematical models composed of coupled reaction-diffusion systems~\cite{Mark70,Scot79,Eilb81,Keen89,Panf90,Panf91,Palm92,Pere95,Taka95,Bose95,Neko99,Binc01,Reut03}  have been used for elucidating the pattern dynamics in these parallel fibers.
Coupled reaction-diffusion systems are also used as models of information processing between neural assemblies~\cite{Babl94,Neko98}. 
In non-biological experiments, it has been reported that mutual synchronization between two chemical waves  occurs in the Belousov-Zhabotinsky chemical reaction system with cross-membrane coupling~\cite{Wins91} and camera-video projector coupling~\cite{Hild03}.  
Understanding and controlling pattern dynamics in coupled reaction-diffusion systems are important research subjects in many  applications. 

In this paper, we report on the pulse dynamics that emerge from a  system of two mutually coupled excitable fibers when the dynamical properties of two excitable fibers are {\it not} identical.
%
Such a situation is not uncommon.
For example, the diameters of real neuronal fibers are generally not equal.
This situation is modeled as a difference of diffusion coefficients for each fiber in a reaction-diffusion system.
%
%
Using the spatially one-dimensional FitzHugh-Nagumo equations~\cite{Fitz61,Nagu62} as a model of a single  excitable fiber, we show that in some cases, two propagating pulses do {\it not} annihilate upon head-on collision and are reconstructed with unchanged spatial profiles like {\it solitons}~\cite{Zabu65}.   
Other interesting and somewhat unexpected pulse dynamics, including {\it recombination of a solitary pulse and synchronized pulses} and {\it overtaking of a slow-moving solitary pulse by fast-moving synchronized pulses}, are also shown.
To our knowledge, these pulse dynamics have not yet been reported in previous  studies on coupled reaction-diffusion systems. 

This paper is organized as follows.
A brief description of the mathematical model for coupled excitable fibers is given in Sec.~2.
%
Section~3 introduces the reentrant wave that is well-known pattern dynamics observed in coupled reaction-diffusion systems, as well as the soliton-like pulse collision that we first report in the present paper.
Section~4 discusses  the stability of the synchronized pulses as the difference between the intra-diffusion coefficients of the two fibers changes.  
%
Section~5 provides several examples of pulse dynamics associated with the destruction of the synchronized pulses. 
Section~6 is a summary and discusses the importance of our findings in the context of neuroscience. 

\section{Model}
We consider two mutually coupled one-dimensional FitzHugh-Nagumo (FHN) fibers.
The system consists of the following equations:  
\begin{eqnarray}
\label{fiber_1}
\begin{split}
\left \{
\begin{array}{ll}
\dot{u}_1 = u_1(u_1-\alpha)(1-u_1) - v_1 + \kappa_1 \nabla^2 u_1 + \epsilon (u_2 - u_1)\\
\dot{v}_1 =  \tau (u_1-\gamma v_1),
\end{array}
\right .
\\
 \left \{
\begin{array}{ll}
\dot{u}_2 = u_2(u_2-\alpha)(1-u_2) - v_2 + \kappa_2 \nabla^2 u_2 + \epsilon (u_1 - u_2) \\
\dot{v}_2 =  \tau (u_2-\gamma v_2).
\end{array}
\right .
\end{split}
\end{eqnarray}
%
Subscripts ``1" and ``2" denote the first and the second fibers.
The state variables $u_{1, 2} = u_{1,2} (x, t)$ and  $v_{1, 2} = v_{1,2} (x, t)$, where $x\in [0, L]$ and $t\in [0, \infty)$ are space and time coordinates, are the activators (membrane potentials) and the inhibitors (recovery variables), respectively.
The parameters of the reaction kinetics are fixed as  $\alpha= 10^{-1}, \tau= 2\times 10^{-3}$, and $\gamma= 2.5$ so that a local kinetics  shows an excitable property; i.e., a small but finite perturbation to the resting state $(u, v)=(0, 0)$ leads to a large excursion. 
The terms  $\nabla^2 u_{1,2} = \partial^2 u_{1,2}/\partial x^2$ represent intra-fiber diffusions and $\kappa_1$ and $\kappa_2$ are their coefficients. 
The value of $\kappa_1$ is fixed at $0.25$ throughout this paper.
 The mutual interaction between two excitable fibers is introduced as linear coupling terms $\epsilon (u_{1,2} - u_{2,1})$ for activators. 
We take $\epsilon$ and $\kappa_2$ to be the control parameters.   

The numerical simulations use the Euler integration scheme with a time step $\Delta t= 10^{-2}$. 
Diffusion terms  at a spatial point $x_i (=i\Delta x)$ with  a spatial step $\Delta x = 5\times 10^{-1}$ are approximated as $\nabla^2 u_{1, 2} (x_i) = (1/(\Delta x)^2) (u_{1,2} (x_{i-1}) - 2 u_{1,2} (x_i) + u_{1,2} (x_{i+1}))$.  
Periodic boundary conditions are employed for both fibers: $u_{1, 2} (0, t) = u_{1, 2} (L,t)$ and   $v_{1, 2} (0, t) = v_{1, 2} (L,t)$.

\section{Reentrant Wave and Soliton-like Pulse Collision}
Here, we investigate the pulse dynamics of the system of Eqs.~(\ref{fiber_1}) when a pulse propagating to the right  is initiated on  fiber 1, and fiber 2 is in the global resting state.
To prepare these states, we consider the following initial conditions: 
\begin{eqnarray}
\label{init_1}
u_1 (x, 0) & = &
\left \{
\begin{array}{lll}
0 \quad {\rm for} \quad 0\le x\le 0.48L \\
1 \quad {\rm for} \quad 0.48 L < x < 0.52 L \\
0 \quad {\rm for} \quad 0.52 L \le x < L,
\end{array}
\right .
\\
v_1 (x, 0) & = & 
\left \{
\begin{array}{lll}
0.1 \quad  {\rm for} \quad 0\le x\le 0.48L \\
\\
0 \ \quad   {\rm for} \quad 0.48 L \le x < L,
\end{array}
\right .
\end{eqnarray}
for fiber 1,  and 
\begin{eqnarray}
\label{init_2}
u_2(x, 0) = v_2 (x, 0) = 0 \quad  \forall x\in [0, L],
\end{eqnarray}
for fiber 2. 
%
Applying these initial conditions and setting the interaction between fibers $\epsilon$ to zero for $t<t_0$, where $t_0$ is a short interval, we obtain the stationary state mentioned above.
We redefine these required states as initial conditions, and interaction between fibers is taken into account.  
When $\kappa_1$ and $\kappa_2$ are identical, we observe the following four different phases after the initial transient dies out with the increase of inter-fiber coupling strength $\epsilon$: (i) a solitary pulse propagating in fiber 1, (ii) formation of the reentrant wave, (iii) the global resting state after a finite repetition of reentrant waves, and (iv) synchronized pulses propagating in both fibers. 
In addition to the these phases,  when $\kappa_1$ and $\kappa_2$ are not equal, a soliton-like collision occurs  depending on the control parameters. 
%
The following sections deal with these observations in detail.  
\subsection{Identical Case}
First, we show what happens  when $\kappa_1$ and $\kappa_2$ are identical;  i.e., $\kappa_1 = \kappa_2= 0.25$.
%

%

\begin{figure}[!t]
\begin{center}
\includegraphics[width=14cm, clip]{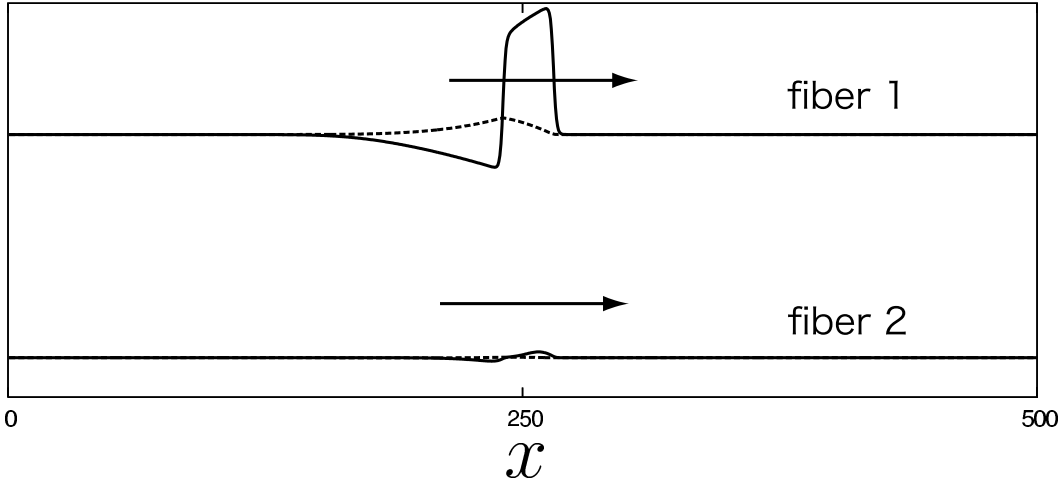}
\caption{Profile of a solitary pulse in fiber 1 with $\epsilon= 5\times 10^{-3}$. 
A tiny sub-threshold excitation is also induced in fiber 2 through the inter-fiber interaction.
Arrows indicate the direction of propagation.
}
\label{figure1}
\end{center}
\end{figure}

As shown in Fig.~\ref{figure1}, for a sufficiently small value of $\epsilon$, a propagating pulse in  fiber 1 does not significantly affect  fiber 2, and only a sub-threshold excitation typically appears as a small amplitude pulse in fiber 2. 

\begin{figure}[!t]
\begin{center}
\includegraphics[width=14cm, clip]{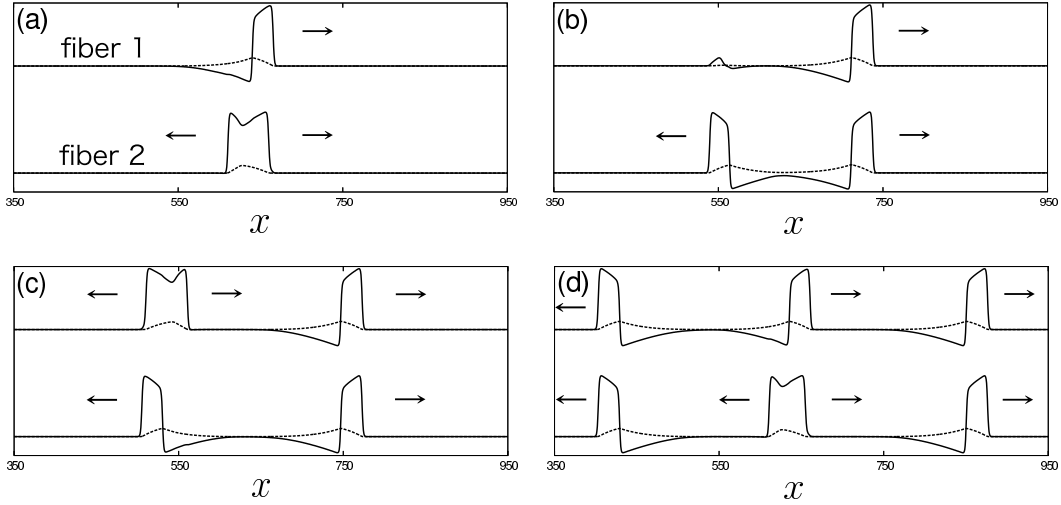}
\caption{A series of snapshots (a) to (d) for the reentrant wave with $\epsilon=7.3\times 10^{-3}$. $t= 180,  450, 580$, and $960$. 
(a) A supra-threshold excitation is induced in fiber 2 by a right-propagating pulse in fiber 1, and it splits into two pulses propagating in opposite  directions. 
(b) The right-propagating pulse in fiber 2 is synchronized with that in fiber 1, whereas the left-propagating pulse remains alone.
(c) The left-propagating pulse in fiber 2 induces a new supra-threshold excitation in fiber 1, and it  splits into two pulses propagating in opposite  directions.
(d) In the same way, the right-propagating pulse in fiber 1 induces   a supra-threshold excitation in fiber 2.
This alternative excitation in the two fibers is the source of the reentrant wave.
}
\label{figure2}
\end{center}
\end{figure}

With increasing $\epsilon$, the qualitative features of the dynamics change dramatically.
%
%
When $\epsilon_{c1} \sim 7.2058\times 10^{-3}$, a propagating pulse in fiber 1 induces a supra-threshold excitation in fiber 2.
Figure~\ref{figure2} shows a series of spatial profiles with $\epsilon = 7.3\times 10^{-3}$.
The new excitation in fiber 2 splits into two pulses moving leftward and rightward as shown in Fig.~\ref{figure2} (a).
The right-propagating pulse in fiber 2 immediately becomes synchronized with the one in fiber 1,
whereas the left-propagating pulse in  fiber 2 remains solitary (Fig.~\ref{figure2} (b)). 
%
Because there is a refractory region behind the right-propagating pulse in fiber 1, a time interval is needed to develop the subsequent  excitation in  fiber 1 induced by the left-propagating pulse in fiber 2 (Fig.~\ref{figure2} (c)). 
%
Consequently, two pulses propagating in opposite directions emerge  from this subsequent excitation in fiber 1, and this new left-propagating pulse is synchronized with the previously generated pulse  in fiber 2, while the right-propagating one is alone for a time. 
This right-propagating pulse will eventually cause another excitation in fiber 2 as shown in Fig.~\ref{figure2} (d).
These processes repetitively occur   in a specific region.
The dynamical pattern associated with such repetitions is called the {\it reentrant wave}~\cite{Panf90,Panf91,Palm92,Pere95}.
The reentrant wave  is considered to be an origin of fatal heart diseases  such as tachycardias and fibrillation~~\cite{Alle73,Krin78}. 
%
%

\begin{figure}[!t]
\begin{center}
\includegraphics[width=14cm, clip]{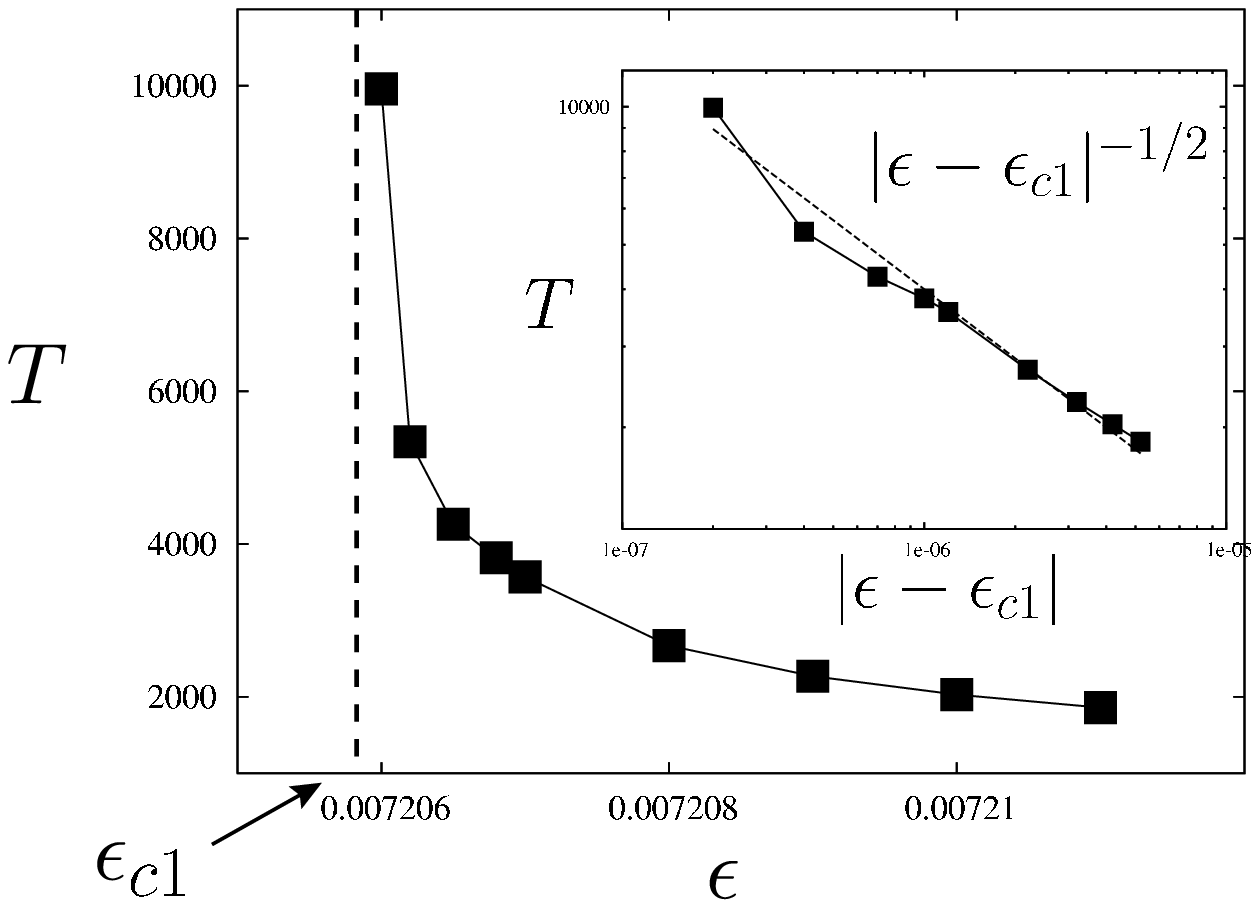}
\caption{Characteristic period $T$ of the reentrant wave as a function of inter-fiber coupling strength $\epsilon$. Inset: The same plot on a logarithmic scale. L=$10^3$. The broken line in the inset indicates $|\epsilon - \epsilon_{c1}|^{-1/2}$.}
\label{figure3}
\end{center}
\end{figure}

%
%
In the reentrant wave phase, alternate generation of pulses repeats  with a characteristic period $T$. 
This characteristic period is depicted in Fig.~\ref{figure3} as a function of the coupling strength.
The period shows a power law with exponent $1/2$, i.e.,  $T\sim |\epsilon - \epsilon_{c1}|^{-1/2}$ near the transition point $\epsilon_{c1}$. 
This result implies that the saddle-node bifurcation is the  onset mechanism of the reentrant wave. 

\begin{figure}[!t]
\begin{center}
\includegraphics[width=14cm, clip]{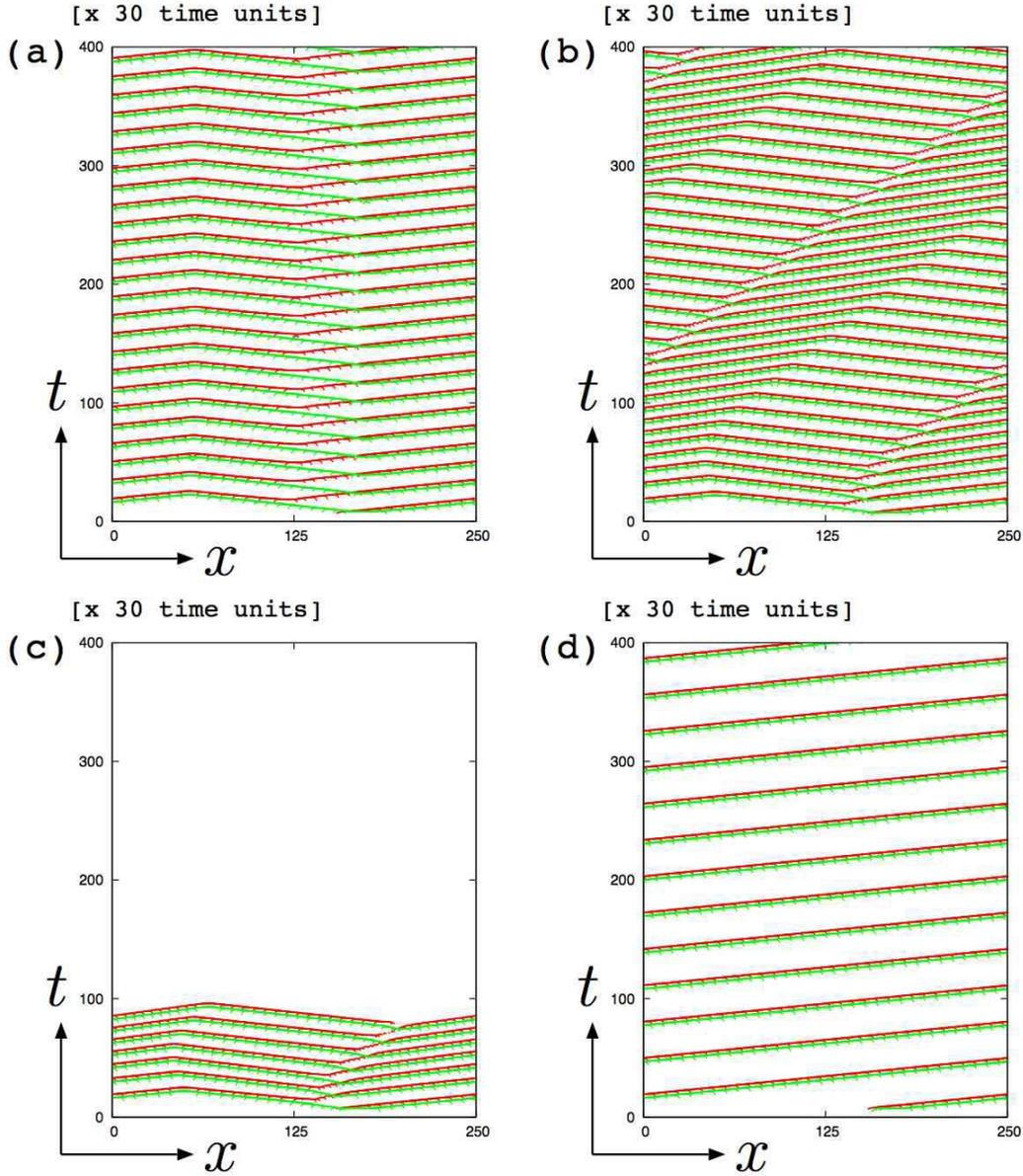}
\caption{Spatio-temporal plots for (a) the reentrant wave: $\epsilon=1.0\times 10^{-2}$, (b) the reentrant wave with drift: $\epsilon = 2.8\times 10^{-2}$, (c) the transient reentrant wave: $\epsilon = 3\times 10^{-2}$, and (d) synchronized pulses: $\epsilon = 10^{-1}$. The red lines  indicate the pulses locations on fiber 1, which is estimated with the criterion that $u_1$ is above $0.7$. The green lines indicate the pulse locations on fiber 2. $L=250$.}
\label{figure4}
\end{center}
\end{figure}

The reentrant wave has  two qualitatively different features depending on the inter-fiber coupling strength $\epsilon$. 
1) For smaller values of $\epsilon$, the core position of the reentrant wave is fixed in time as shown in Fig.~\ref{figure4} (a).
2) As the coupling strength becomes larger, the core position of the reentrant wave starts to drift to the right. 
A spatio-temporal plot for the drifted reentrant wave is shown in Fig.~\ref{figure4} (b). 
The direction of drift depends on the initial conditions, and is the same as  that of the propagating pulse in fiber 1. 
%
%
%
%
%

For $\epsilon > \epsilon_{c2} \sim 2.83\times 10^{-2}$, the reentrant wave disappears after a finite repetition as shown in Fig.~\ref{figure4} (c).
The lifetime  of the transient reentrant wave decreases as the value of $\epsilon - \epsilon_{c2}$ increases. 
%
%
%
%
When a solitary pulse propagates in a fiber, there is a zone in the other fiber that is stimulated by interacting with the pulse.
This excitatory induction is stronger for larger $\epsilon$.
Thus, the time interval required to generate the new pulse becomes shorter, and the period of alternate pulse generation decreases.
Furthermore, for a given value of $\epsilon$, the period is initially longer, and converges to a stationary value that is smaller than the initial period.
In other words, the distance between the points at which new excitations emerge decreases over time.
For $\epsilon_{c2}<\epsilon<\epsilon_{c3}$, a new excitation at one time does not split into two propagating pulses because the half side of a new excitation meets the refractory region.
The repetition subsequently terminates.

Finally, for sufficiently large values of $\epsilon > \epsilon_{c3} \sim 7.2\times 10^{-2}$, the initial pulse in fiber 1 generates a new excitation in fiber 2.
However, in this case, the new excitation does not form two pulses propagating in opposite directions.
Only the right-propagating pulse emerges from the excitation and immediately synchronizes with the initially generated pulse in fiber 1(Fig.~\ref{figure4} (d)).

\begin{figure}[!t]
\begin{center}
\includegraphics[width=14cm, clip]{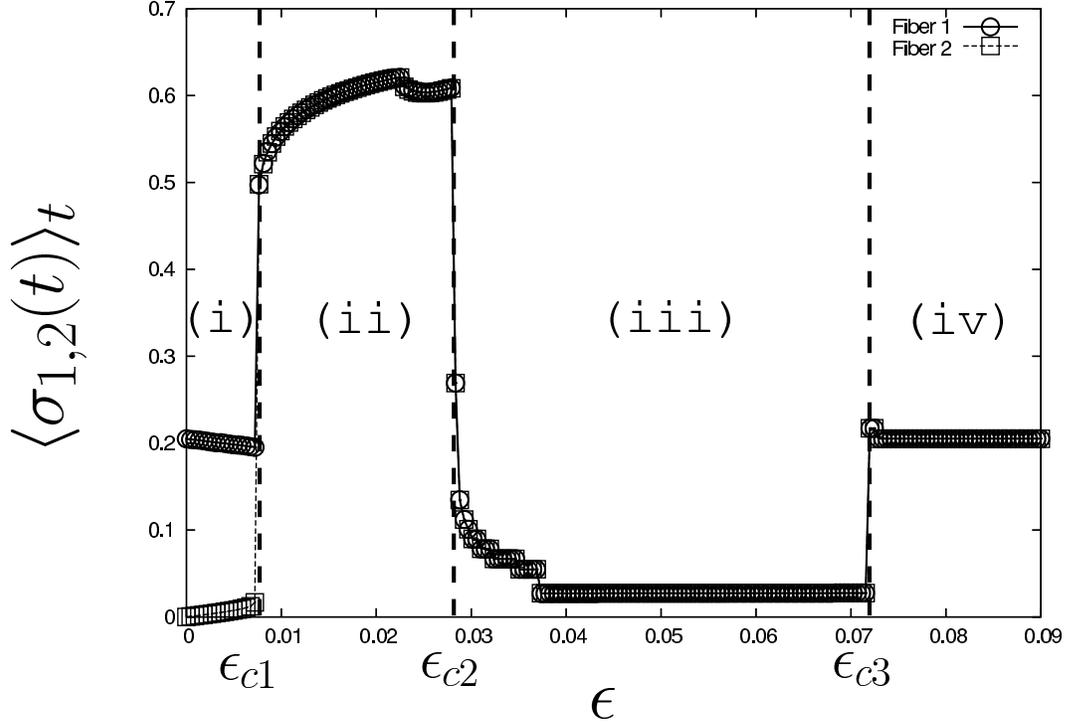}
\caption{Time averaged norms $\langle \sigma_{1,2} (t)\rangle_t$ vs. $\epsilon$ for the case of $\kappa_1 = \kappa_2$. Stationary states are (i) a solitary pulse in the fiber 1, (ii) the reentrant wave, (iii) the global resting state after the transient reentrant waves, and (iv) synchronized pulses. $L=10^3$.}
\label{figure5}
\end{center}
\end{figure}

%
To characterize these phases, we use the following spatially coarse grained quantity: 
\begin{eqnarray}
\label{sigma}
\sigma_{1, 2} (t) = \sqrt{\frac{1}{L}\int_{0}^{L} {\rm d}x (u_{1,2} (x, t)^2 + v_{1, 2} (x, t)^2)}.
\end{eqnarray}
%
Figure~\ref{figure5} shows the time averages $\langle \sigma_{1, 2} (t) \rangle_t$ as functions of the inter-fiber coupling strength $\epsilon$. 
The four phases of spatio-temporal patterns -- (i) solitary pulse, (ii)  reentrant wave,  (iii)   transient reentrant wave, and (iv)  synchronized pulses -- are clearly distinguished as $\langle\sigma_{1, 2} (t)\rangle_t$ changes.

\subsection{Nonidentical Case}
Next, turn our attention to the case in which $\kappa_1$ and $\kappa_2$ are not identical.
This subsection includes  new findings which have not been reported in previous  studies on coupled reaction-diffusion media.
In the following, we set $\kappa_1$ to $0.25$ and $\kappa_2$ to $0.09$, and focus on the pulse dynamics in fiber 2.

As described in the identical case, when  $\epsilon$ is too small, 
a sub-threshold excitation in fiber 2 is generated under the influence of the pulse propagating in fiber 1. 
%
The solitary pulse does not significantly affect the other fiber.

However, when $\epsilon$ exceeds $\epsilon_{c'1}\sim 7.05 \times10^{-3}$, an excitation  in  fiber 2 is induced by the propagating pulse in fiber 1.
%
%
This excitation generates two pulses propagating in opposite directions in fiber 2.
%
%
This situation occurs in the same manner as described in the identical fibers case. 
%
For non-identical fibers, asymmetrical excitations, which we call {\it one-way excitations}, appear in the following way.
Through inter-fiber interactions, the left-propagating pulse in  fiber 2 activates a localized zone in fiber 1, which corresponds to the action potential region in fiber 2.
%
%
%
%
%
Because diffusion generally acts as a smoother for a given spatial inhomogeneity, the activated zone in fiber 1 can not be excited due to the larger diffusion coefficient ($\kappa_1=0.25$).
On the other hand, the activated zone in fiber 2 is easily excited to form a pulse because of  the smaller coefficient ($\kappa_2=0.09$), resulting in the one-way excitation.
Indeed, compared with the case of identical fibers, a stronger intra-fiber coupling is needed to form the reentrant wave.
The one-way excitation is one of the new phases that emerge through the interaction between non-identical fibers, and it exists between the solitary pulse and the reentrant wave phases. 
Furthermore, we found the following interesting pulse dynamics in this phase.

\begin{figure}[!t]
\begin{center}
\includegraphics[width=14cm, clip]{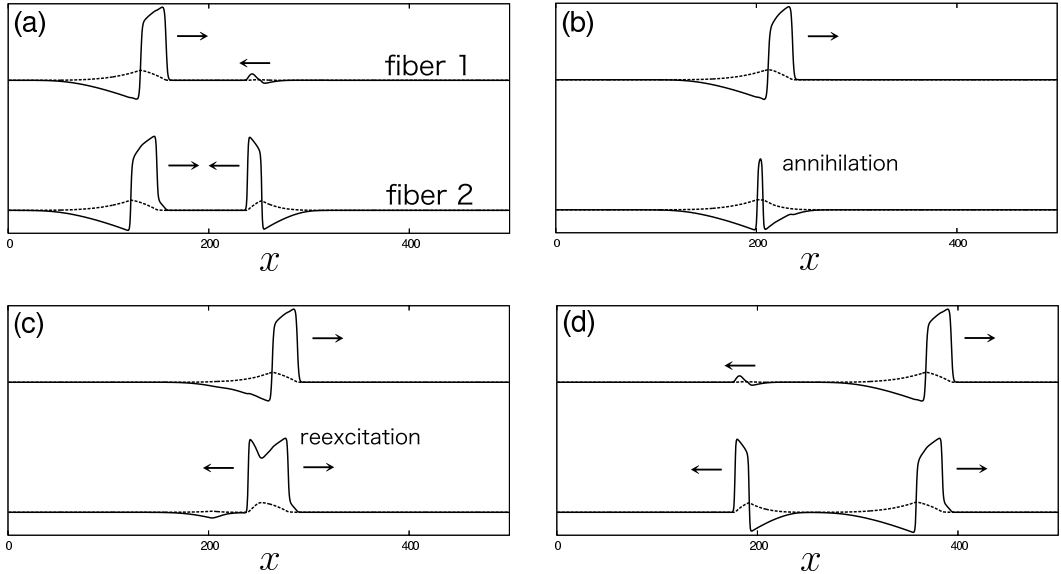}
\caption{A series of snapshots (a) to (d) for the soliton-like pulse collision with $\kappa_1 = 0.25, \kappa_2 = 0.09$, and $\epsilon=8\times 10^{-3}$. $t= 1.0\times 10^3,  1.3\times 10^3, 1.5\times 10^3$, and $1.9\times 10^3$.
(a) Head-on collisions occur in both fibers. 
(b) Mutual annihilation occurs in fiber 2 whereas the pulse prevails over the sub-threshold excitation in  fiber 1.
(c) A supra-threshold excitation that splits  into two propagating pulses is induced in fiber 2 by the pulse in fiber 1.
(d) The spatial profiles of all pulses recover after head-on collisions.}
\label{figure6}
\end{center}
\end{figure}

\begin{figure}[!t]
\begin{center}
\includegraphics[width=14cm, clip]{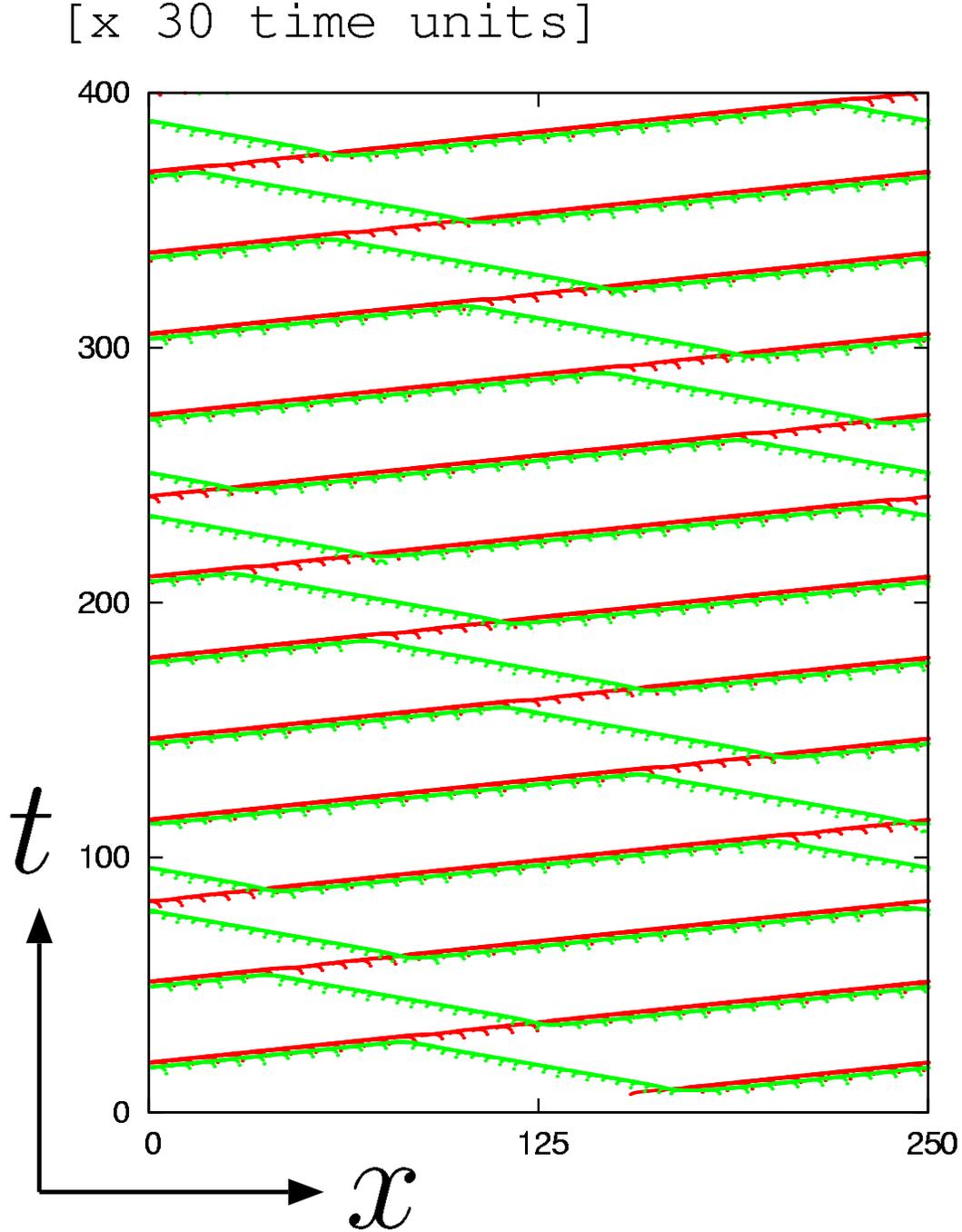}
\caption{Spatio-temporal plot for a soliton-like pulse collision  with $\kappa_1 = 0.25, \kappa_2 = 0.09$, and $\epsilon=8\times 10^{-3}$. $L=250$.}
\label{figure7}
\end{center}
\end{figure}

Figure~\ref{figure6} shows typical snapshots of spatial profiles in the  one-way excitation phase.  
Due to  the periodic boundary condition, two pairs of pulses propagating in opposite directions become close to each other (Fig.~\ref{figure6} (a));  i.e., one pair is composed of  synchronized pulses, which are two supra-threshold excitations both in fibers 1 and 2 with a tiny delay, and the other pair is composed of a solitary pulse propagating in fiber 2 and a sub-theshold excitation in  fiber 1. 
Eventually,  head-on collisions occur in both fibers (Fig.~\ref{figure6} (b)). 
In fiber 1, the head-on collision between the supra-threshold  pulse and the sub-threshold excitation does not lead to mutual annihilation; i.e., the sub-threshold excitation vanishes  and the right-propagating pulse persists.
On the other hand, in fiber 2, the two propagating pulses mutually annihilate once by the head-on collision. 
A new supra-threshold excitation, however,  is regenerated through induction from  the right-propagating pulse in fiber 1.
This excitation splits into two pulses propagating in opposite directions (Fig.~\ref{figure6} (c)).
Finally, the spatial profiles of all pulses completely recover   after the head-on collision (Fig.~\ref{figure6} (d)).
These kinematic features are similar to the characteristics  of {\it solitons}~\cite{Zabu65} in integrable systems. 
A spatio-temporal plot of this solition-like collision is also shown in Fig.~\ref{figure7}.

Theoretical studies have predicted that head-on collisions between propagating pulses do not always lead to mutual annihilation, but form a variety of nontrivial dynamical behavior, including soliton-like collisions~\cite{Tuck78,Bar92,Petr94,Kris94,Kose95,Ohta97,Arge97,Nish03}, and there are experiments 
that verify their  existence~\cite{Rote91,Will92}. 
%

%
%
%
%
It is also known that a transition from annihilation to crossing of the pulses upon a head-on collision occurs in an FHN system~\cite{Arge00} that considers diffusion of the inhibitor $v$ and bistable reaction kinetics; i.e., coexistence of a stable limit cycle and a stable fixed point.
In real nerve systems, however, diffusion for the inhibitor generally does not exist, because the inhibitor represents potential-dependent gating variables, and a fiber usually has a simple excitable property.
In our system, we do not require both diffusion for the inhibitor and multi-stable kinetics. 
Instead, fiber 1 plays the role of an auxiliary field for the occurrence of the soliton-like behavior in fiber 2.

\begin{figure}[!t]
\begin{center}
\includegraphics[width=14cm, clip]{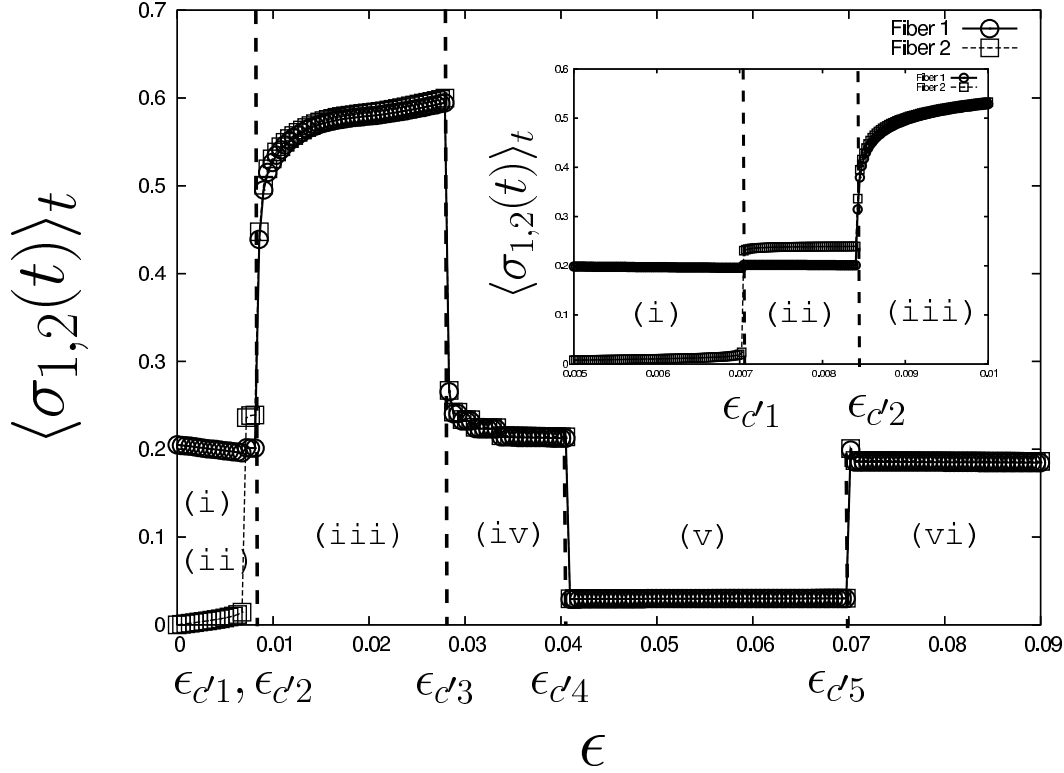}
\caption{Time averaged norms  $\langle \sigma_{1,2} (t)\rangle_t$ vs. $\epsilon$ for the case of $\kappa_1 \neq \kappa_2$.  Inset: An enlargement of the region $\epsilon \in [5\times 10^{-3}, 10^{-2}]$. Six different phases exist in the case of coupled non-identical fibers: (i) a solitary pulse in fiber 1, (ii) a soliton-like pulse collision, (iii) the reentrant wave, (iv) synchronized pulses after the transient reentrant waves, (v) the global resting state after the transient reentrant waves, and (vi) synchronized pulses. $\kappa_1 =0.25, \kappa_2 = 0.09$  ($\Delta\kappa = 0.16$), and  $L=10^3$.}
\label{figure8}
\end{center}
\end{figure}

\begin{figure}[!t]
\begin{center}
\includegraphics[width=14cm, clip]{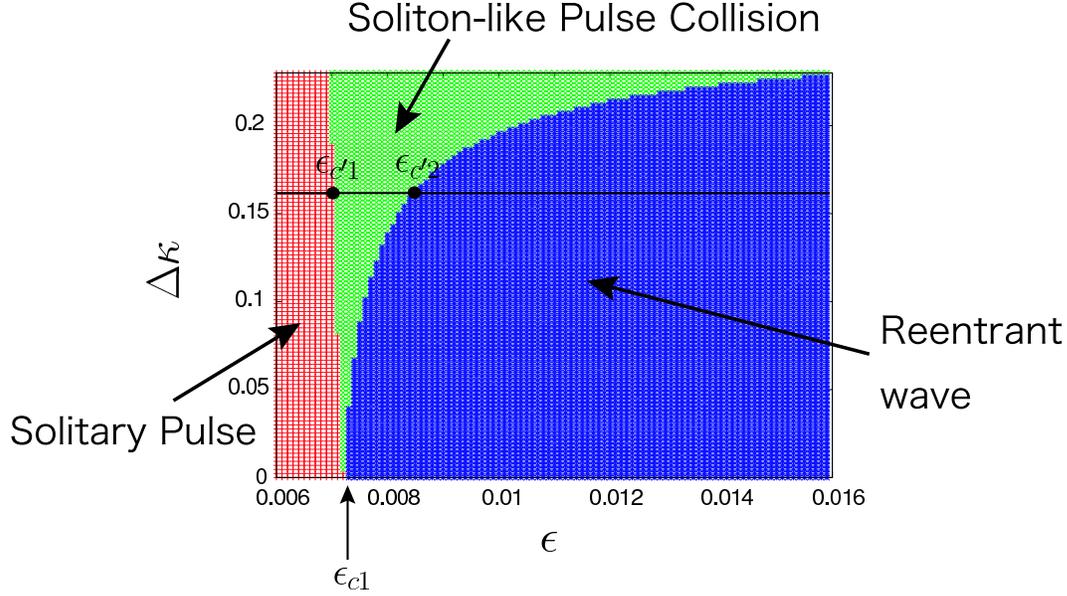}
\caption{Phase diagram in the plane of $(\epsilon, \Delta \kappa), \Delta \kappa = \kappa_1 - \kappa_2$. $\kappa_1$ is set to $0.25$ and $L=10^3$.}
\label{figure9}
\end{center}
\end{figure}

 As $\epsilon$ increases, the reentrant wave phase and the synchronized pulses phase appear in non-identical fibers, as well as in the identical fibers.
The phase diagram of the nonidentical case is depicted  in Fig.~\ref{figure8}.
%
For $\epsilon>\epsilon_{c'2}$, i.e., after the formation of reentrant wave,  the qualitative behaviors are similar to those of the indentical fibers as shown in Fig.~\ref{figure5}.
However,  a new phase appears, as indicated by (iv) in Fig.~\ref{figure8}.
In this phase, synchronized pulses form after a finite repetition of the reentrant waves.  
%
We also  investigated the phase diagram in two-dimensional parameter space $(\epsilon, \Delta \kappa)$, where $\Delta\kappa = \kappa_1 - \kappa_2$.
The result is shown in Figure~\ref{figure9} . 
There are clear bifurcation curves that discriminate different phases.
 The soliton-like phase still survives for tiny $\Delta \kappa$, and the interval of $\epsilon$ for it extends as $\Delta \kappa$ increases. 
 %

\section{Stability of Synchronized Pulses}
In the previous section, we considered the case where a right-propagating pulse is initiated only in fiber 1 at $t=0$. 
As shown in Figs.~\ref{figure5} and \ref{figure8}, 
for such initial conditions,  synchronized pulses are observed for $\epsilon\gtrsim \epsilon_{c3}$ in the identical case and $[\epsilon \in (\epsilon_{c'3}, \epsilon_{c'4})] \vee [\epsilon\gtrsim \epsilon_{c'5}]$ in the nonidentical case.  
It is expected, however,  that when $\Delta \kappa = \kappa_1 - \kappa_2$ is small, synchronized pulses will be in a stable state for small $\epsilon$ if a pair of synchronized pulses are taken as initial conditions.   
From the viewpoint of information processing in the brain, the stability of synchronized propagating pulses in a nerve bundle is of interest. 
Thus, in this section, we consider the stability of the synchronized pulses against changes in $\Delta \kappa$ and $\epsilon$ when a pair of synchronized pulses are initiated at $t=0$. 
$\kappa_1$ is set to $0.25$ and $\kappa_2$ is varied within an interval $[0, 0.25]$ in the following. 
To detect whether the synchronization is stable or not, we consider the  velocities $c_1$ and $c_2$ of the two pulses propagating in  fiber 1 and 2, respectively. 

\begin{figure}[!t]
\begin{center}
\includegraphics[width=14cm, clip]{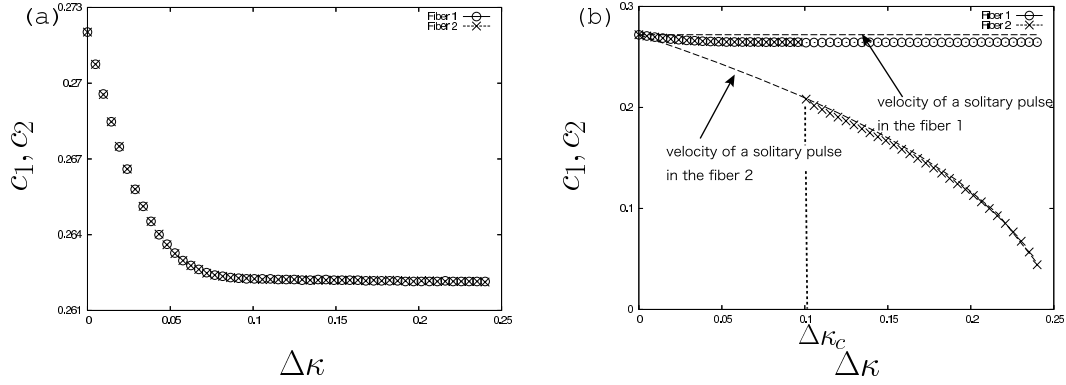}
\caption{Velocities $c_1$ and $c_2$ of  propagating pulses in fibers 1 and 2 as  functions of $\Delta \kappa = \kappa_1 - \kappa_2$ with  (a) $\epsilon = 8\times 10^{-3}$, and (b) $\epsilon = 6\times 10^{-3}$. Open circles indicate $c_1$ and crosses $c_2$. $\kappa_1$ is set to $0.25$. Velocities of solitary pulses in the two fibers without the inter-fiber interaction (i.e., $\epsilon = 0$) are  plotted as broken lines in (b). 
}
\label{figure10}
\end{center}
\end{figure}

Figure~\ref{figure10} (a) plots $c_1$ and $c_2$ as functions of $\Delta \kappa$ for a relatively larger $\epsilon$($= 8\times 10^{-3}$). 
Although both velocities decrease with increasing $\Delta \kappa$, the result of $c_1 = c_2$ for any $\Delta \kappa$ indicates that two propagating pulses remain synchronized. 
On the other hand, for smaller $\epsilon$($= 6\times 10^{-3}$), there is a sudden dip in the graph of $c_2$ vs. $\Delta \kappa$ at $\Delta \kappa_c \sim 0.1$ whereas the graph of $c_1$ vs. $\Delta \kappa$ is continuous, as shown in Fig.~\ref{figure10} (b). 
This result indicates that the synchronization becomes unstable for $\Delta\kappa > \Delta\kappa_c$.
Indeed,  even if synchronized pulses are initially given, the pulses become de-synchronized and propagate with different velocities.  

\begin{figure}[!t]
\begin{center}
\includegraphics[width=14cm, clip]{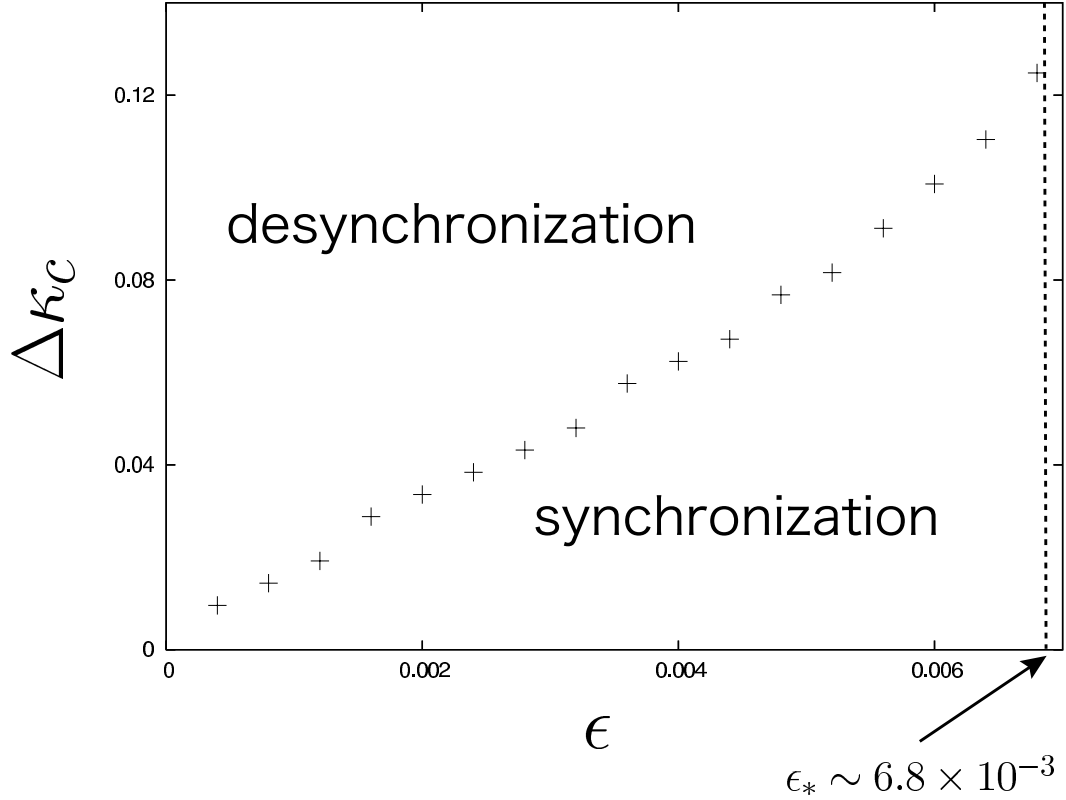}
\caption{Critical difference $\Delta \kappa_c$ between intra-diffusion coefficients as a function of $\epsilon$. $\kappa_1$ is set to $0.25$.
Beyond the critical value, the synchronous pulses become unstable, and split into two solitary pulses with different propagating speeds.
}
\label{figure11}
\end{center}
\end{figure}

The dependence of the critical difference $\Delta \kappa_c$ on $\epsilon$ is shown in Figure~\ref{figure11}. 
%
There is a critical inter-fiber coupling strength $\epsilon_* \sim 6.8\times 10^{-3}$, beyond which the synchronization is sustained for any $\Delta\kappa\le 0.25$.  

\section{Recombination and Overtaking}
\begin{figure}[!t]
\begin{center}
\includegraphics[width=14cm, clip]{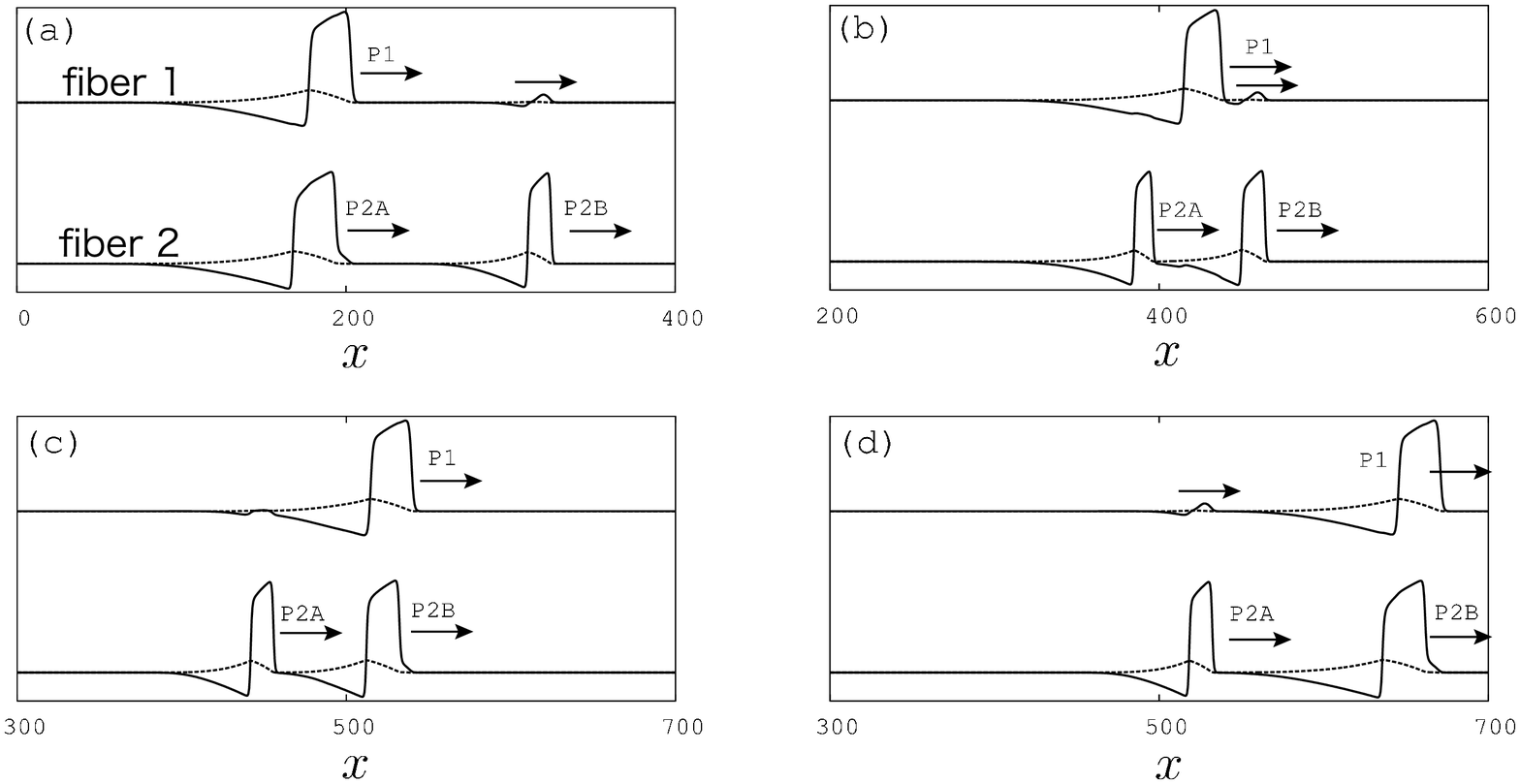}
\caption{A series of snapshots (a) to (d) for recombination of synchronized and solitary pulses with $\kappa_1 = 0.25, \kappa_2 = 0.09$, and $\epsilon = 8\times 10^{-3}$. $L=700$. 
$t=4.8\times 10^3, 5.7\times 10^3, 6.0\times 10^3$, and $6.6\times 10^3$. 
(a) Synchronized pulses composed of P1 and P2A comes close to P2B.
(b) Synchronization is broken by the inhibitor behind P2B.
(c) New synchronized pulses  composed of P1 and P2B form.
(d) After recombination, the new synchronized pulses move away from P2A.}
\label{figure12}
\end{center}
\end{figure}

%
%
In the soliton-like phase, two pulses in fiber 2 face each other as shown in Fig.~\ref{figure6} (a).
%
%
Here, we consider an initial state such that there are two pulses propagating in the {\it same} direction in fiber 2; one is synchronized with  a pulse in fiber 1, and the other is a solitary pulse as depicted in Fig.~\ref{figure12} (a). 
%
This initial state is also established by an appropriate initial stimulation at the same parameter values for which the soliton-like collision is observed. 
Here, we set $\kappa_1 = 0.25, \kappa_2 = 0.09$, and $\epsilon=8\times 10^{-3}$.
Starting from the initial state (Fig.~\ref{figure12} (a)), we observe interesting dynamical behavior associated with the loss of  synchronized pulses.
%
%
%

\begin{figure}[!t]
\begin{center}
\includegraphics[width=14cm, clip]{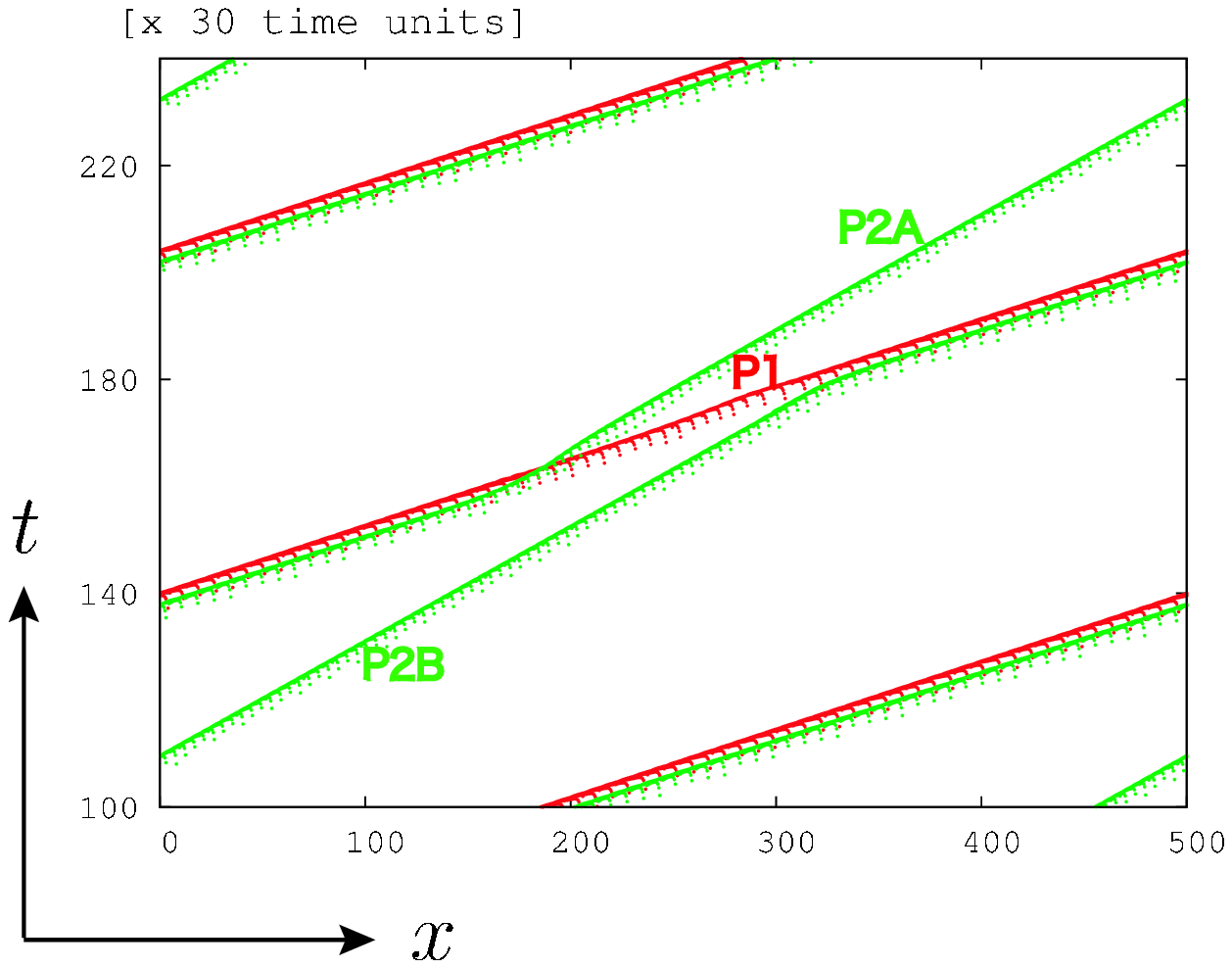}
\caption{Spatio-temporal  plot for the recombination of synchronized and solitary pulses with $\kappa_1 = 0.25, \kappa_2 = 0.09$, and $\epsilon = 8\times 10^{-3}$. $L=500$.
Snapshots are also shown in Fig.~\ref{figure12}.
}
\label{figure13}
\end{center}
\end{figure}

%
In the following, we use the notation P1, P2A, and P2B to indicate the three excited pulses in Fig.~\ref{figure12} (a). 
All three pulses propagate to the right,  and P1 and P2A are synchronized with each other.  
As shown in Fig.~\ref{figure10} (b), 
the synchronized pulses are faster than the solitary pulse in fiber 2.
Therefore, P2A is faster than P2B, and the distance between P2A and P2B decreases  with time.
However, when P2A comes sufficiently close to P2B, the highly concentrated region of  the inhibitor behind P2B decelerates P2A, leading to loss of synchronization between P1 and P2A (Fig.~\ref{figure12} (b)).
%
%
Since P1 is faster than P2A and P2B, synchronization between P1 and P2B is eventually achieved (Fig.~\ref{figure12} (c)).
These new synchronized pulses move away from P2A (Fig.~\ref{figure12} (d)).
We call this series of dynamic processes {\it recombination} of a solitary pulse and synchronized pulses. 
A corresponding spatio-temporal plot is shown in Fig.~\ref{figure13}.

\begin{figure}[!t]
\begin{center}
\includegraphics[width=14cm, clip]{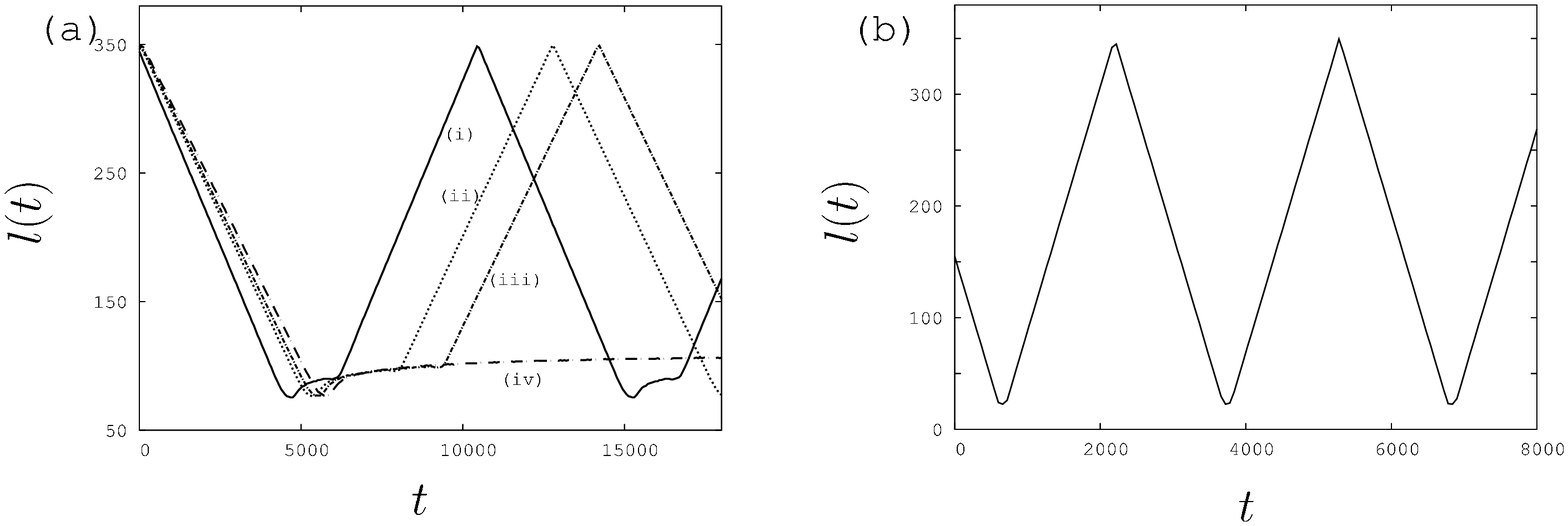}
\caption{(a) Time evolution of the distance $l$ between two pulses P2A and P2B in fiber 2 (see Figs.~\ref{figure12} and \ref{figure13}).  (i) $\Delta\kappa = 0.1$, (ii) $\Delta\kappa=0.09$, (iii) $\Delta\kappa = 0.087$, and (iv) $\Delta\kappa = 0.085$.  
For smaller $\Delta \kappa$, the recombination comes to halt, and a locking state appears (see Fig.~\ref{figure15} for snapshot).
(b) The graph of $l$ vs. $t$ is also shown for $\Delta\kappa = 0.24$.
For larger values of $\Delta \kappa$, the qualitative feature changes; i.e., the behavior of fast-moving  synchronized pulses overtaking a slow-moving solitary pulse emerges (see Fig.~\ref{figure16} for the series of snapshots).
 $\epsilon = 7.3\times 10^{-3}$ and $L=700$.}
\label{figure14}
\end{center}
\end{figure}

To clarify how this recombination depends on $\Delta\kappa$, 
we investigate the time evolution of the distance $l$ between P2A and P2B in fiber 2 for $\epsilon=7.3\times 10^{-3}$.
Depending on the value of $\Delta\kappa$, two qualitatively different  results are obtained, as shown in Figs.~\ref{figure14} (a) and (b). 
%
%

\begin{figure}[!t]
\begin{center}
\includegraphics[width=14cm, clip]{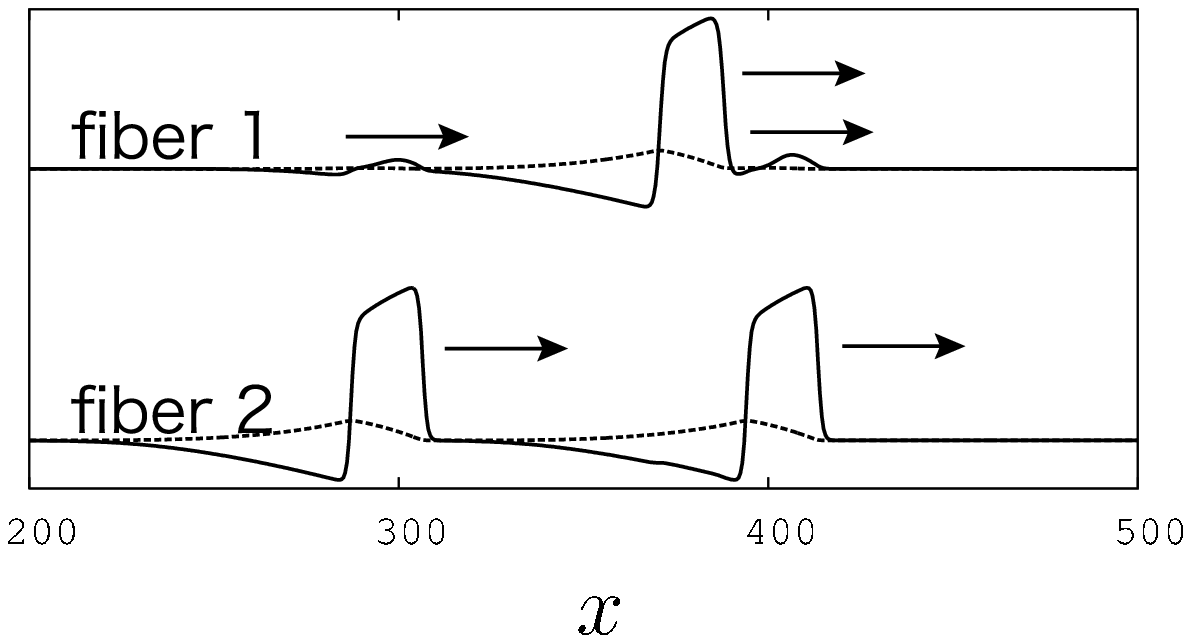}
\caption{A snapshot for the phase locked pulses. $\Delta\kappa = 0.085$ ($\kappa_1 = 0.25$ and $\kappa_2=0.165$) and $\epsilon = 7.3\times 10^{-3}$.
The pulse formation is fixed in time.}
\label{figure15}
\end{center}
\end{figure}

Figure~\ref{figure14} (a) plots $l$ vs. $t$ for four different values of $\Delta\kappa$.
For each graph, the distance between P2A and P2B becomes short over time and approaches a local minimum.   
The time at which $l$ reaches its local minimum corresponds to  the moment when synchronization between P1 and P2A is broken.
After that moment, there is a time interval in which  $l$ slowly increases.
This time interval corresponds to the recombination process.   
Focusing on the change in the graphs from (i) to (iv) in Fig.~\ref{figure14} (a), it is seen that the elapsed  time required for  recombination gradually increases with decreasing  $\Delta\kappa$, but  diverges  at $\Delta\kappa = 8.5\times 10^{-2}$.
%
This result implies that the distance between P2A and P2B converges to a stationary value and the recombination is abandoned along the way as shown in Fig.~\ref{figure15}.
This profile is similar to the one in Fig.~\ref{figure12} (b); however,  all three pulses propagate with the same speed, thus, the relative positions among three pulses are fixed in time.
We call these dynamics  {\it locking} of propagating pulses.

\begin{figure}[!t]
\begin{center}
\includegraphics[width=14cm, clip]{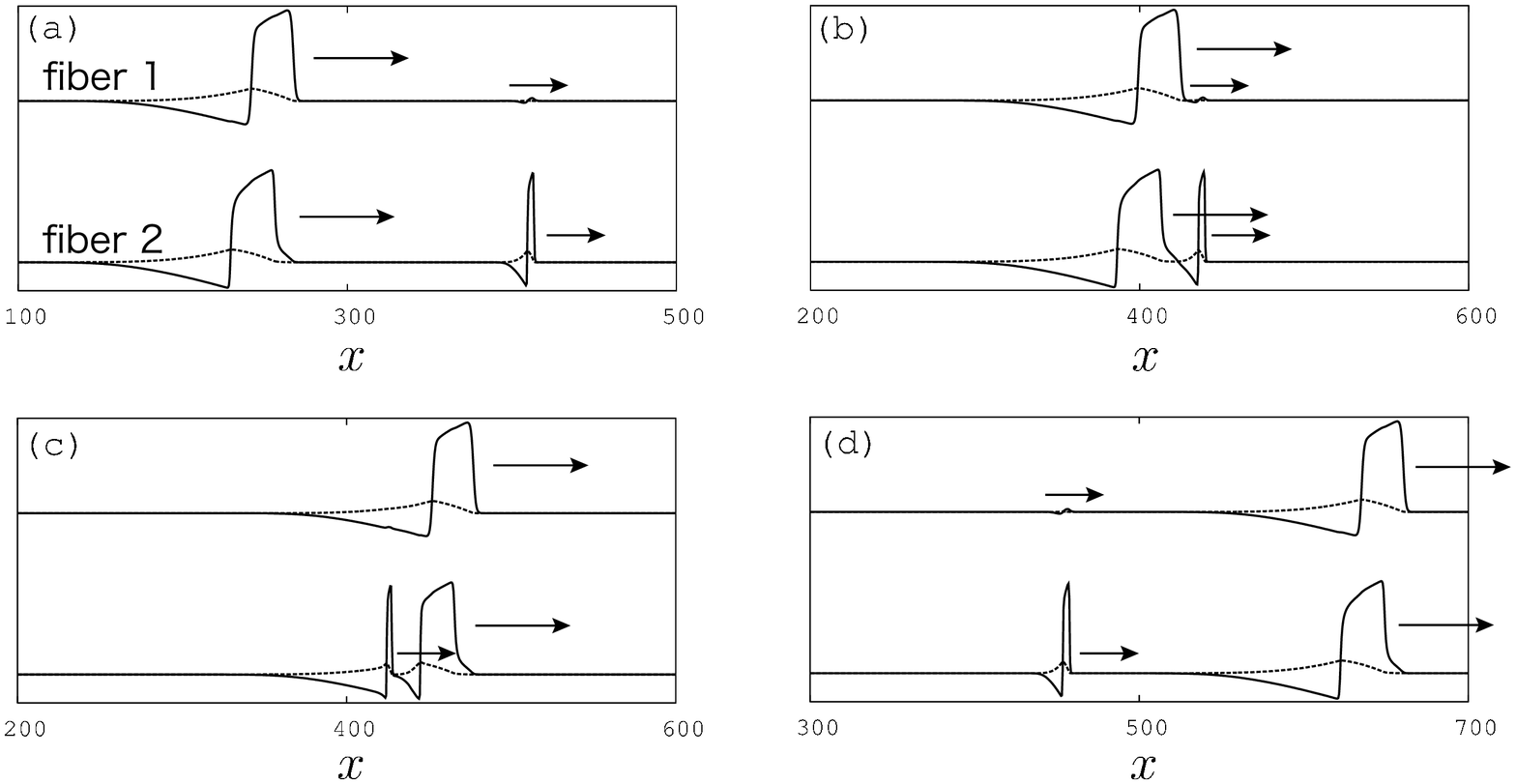}
\caption{A series of snapshots (a) to (d) for fast-moving synchronized pulses overtaking a slow-moving solitary pulse for $\Delta\kappa = 0.24$ ($\kappa_1=0.25$ and $\kappa_2 = 0.01$). $\epsilon = 7.3\times 10^{-3}$, and $L=700$.
$t=2.4\times 10^3, 3.0\times 10^3, 3.2\times 10^3$, and $3.9\times 01^3$.}
\label{figure16}
\end{center}
\end{figure}

For larger values of $\Delta \kappa$, the qualitative feature of the graph, $l$ versus $t$, changes as shown in Fig.~\ref{figure14}  (b); 
i.e., the time interval required for the recombination becomes very short and a clear V-shaped structure appears.
%
Figure~\ref{figure16} shows a series of profiles at  $\Delta\kappa = 0.24$.
%
It seems that the slow-moving solitary pulse is  {\it overtaken} by the fast-moving synchronized pulses. 

\section{Summary and Discussion}
We investigated pulse dynamics in mutually coupled excitable fibers 
%
and showed that new  pulse dynamics occur: {\it solition-like pulse collisions}, whereby  the spatial profiles of pulses completely recover after head-on collisions, {\it recombination}, whereby the combination of synchronized pulses changes into a new one, {\it locking}, whereby  the recombination process is abandoned along the way, and {\it overtaking}, whereby a slow-moving solitary pulse is overtaken by fast-moving synchronized pulses.
%
These exotic behaviors come from the interaction between fibers with different intra-diffusion coefficients.
In other words, one fiber acts as a ``hidden variable" for the other fiber.
Thus,  excitable media, which have FHN-type reaction kinetics, show these exotic behaviors.

Before ending, we should discuss the significance of the present study in the context of neuroscience. 
As mentioned in Sec.~I, there exist densely packed bundles of nerve axons in several regions of the brain. 
%
In such a structure, one can expect an electrical interaction between adjacent nerve axons.
We introduced diffusive couplings $\epsilon (u_{1,2} - u_{2,1})$ in Eqs.~(\ref{fiber_1}) as  electrical interactions between two excitable fibers.
%

In general, however, most nerve axons are  insulated by a lipid material called the {\it myelin sheath} with periodic gaps of exposure called the {\it nodes of Ranvier} in order to  increase  the speed and the reliability of  conduction of pulses~\cite{Keen98}.
Because the myelin sheath reduces current flow that leaks out across the membrane, the electrical interaction between neighboring axons surrounded by myelin sheaths is very weak; hence it is said that the effect of electric coupling between axons is negligible~\cite{Segu86}.
Theoretical studies~\cite{Binc01,Reut03} have also shown that the relative locations of the nodes of Ranvier  on two neighboring myelinated nerve axons influence the degree of synchronization between propagating pulses. 
It would be interesting to extend our study to incorporate a  spatially inhomogeneous structure such as the myelin sheath and the nodes of Ranvier into the system of Eqs.~(\ref{fiber_1}) for actual biological applications.
%
Such a property could be an interesting topic of future work.  

On the other hand, it is also known that there are nerve bundles composed of {\it unmyelinated} axons.
In fact, olfactory nerve axons are unmyelinated and arranged in densely packed bundles~\cite{Boki01}, and the question of  whether there are gap junctional couplings between adjacent axons has been also discussed~\cite{Blin03}. 
We have mainly focused on the pulse dynamics when the diffusion coefficients $\kappa_1$ and $\kappa_2$ in Eqs.~(\ref{fiber_1}) are not equal.
The difference between coefficients might correspond to the difference in the diameters of the two nerve axons.
Such an assumption is quite natural when considering real nerve axons.
Thus, real neural systems should have the ability to produce a wide variety of pulse dynamics similar to the ones shown in the present paper.  

Our results strongly suggest that a nerve axon is not only a cable for transmitting an input signal to the terminal but also a functional element that can process input information by ``crosstalk" between adjacent axons in a bundle~\cite{Deba04}.   
Such crosstalk is not desirable for the purpose of precise conduction of the input pulse, and may be also involved in serious neural diseases such as Multiple Sclerosis (MS), which is caused by the loss of the myelin sheath~\cite{Reut03}.
It is also known that the electrical interaction between lesioned nerve axons is an origin of neuropathic pain such as hyperalgesia and allodynia~\cite{Brid01}.

On the other hand, 
crosstalk among nerve axons may play a constructive role in the brain. 
%
%
%
%
In ~\cite{Boki01}, it is suggested that electrical interactions between neighboring axons in the olfactory nerves contribute to olfactory discrimination by modulating the frequency of the action potentials in neighboring axons and by enhancing the degree of synchronization between their firings.
Such a situation can be also found in mossy fibers in the hippocampus.
The structure of hippocampal mossy fibers is similar to that of the olfactory nerves; i.e., the nerve axons are unmyelinated and arranged in densely packed bundles.
Thus, electrical axo-axonal interactions may occur in mossy fibers and are likely to impact the mechanism of neuronal integration~\cite{Ikeg06}. 
A recent experimental study revealed that depolarization at an axon terminal strongly influences the efficiency of synaptic transmission in mossy fibers~\cite{Alle06}.   
%
In our simple mathematical model, a pulse in a fiber influences the propagation of pulses in the other fibers.
Indeed, synchronization and locking between propagating pulses are established depending on the parameter, i.e., the coupling strength between fibers, and the difference between the  intra-diffusion coefficients.
The interspike interval in the early stage of propagation will change as the pulse travel in fibers.
Thus, synchronization and locking regulate the interspike interval.
This fact implies that synaptic transmission might be affected by electrical axo-axonal interactions in real neuronal systems.

Various types of pulse dynamics caused by axo-axonal interactions in the nerve bundle may act as information processing in the brain~\cite{Deba04}.
The results presented in this paper provide a new insight into the functional aspects of nerve axons. 
 
\section*{Acknowledgments}
We thank T.~Aoyagi for useful suggestions and comments on coupled excitable fibers.
We also thank Y.~Ikegaya for helpful discussions on neuroscience, and K.~Mabuchi and Y.~Nishiura for informing us about useful technical references.
This study is partially supported by Grant-in-Aid No. 17022012 for
Scientific Research on Priority Areas System study on higher-order brain
functions from the Ministry of Education, Culture, Sports, Science and
Technology of Japan.

\end{document}